\documentclass[conference]{IEEEtran}
\IEEEoverridecommandlockouts
\usepackage{amsmath,amssymb,amsfonts}
\usepackage{algorithmic}
\usepackage{graphicx}
\usepackage{textcomp}
\usepackage{xcolor}
\def\BibTeX{{\rm B\kern-.05em{\sc i\kern-.025em b}\kern-.08em
    T\kern-.1667em\lower.7ex\hbox{E}\kern-.125emX}}

\newtheorem{definition}{Definition}

\usepackage{braket}
\DeclareMathOperator{\Tr}{Tr}
\usepackage{bm}
\usepackage{tikz}
\usetikzlibrary{quantikz}
    
\usepackage[style=ieee,backend=bibtex, minnames=3, maxnames=3]{biblatex}
\DeclareFieldFormat{title}{#1\isdot}
\DeclareFieldFormat{journaltitle}{#1\isdot}
\bibliography{bibliography}    

\begin{document}

\title{Quantum variational learning for entanglement witnessing
\thanks{Accepted for publication at IEEE WCCI 2022.}
}

\author{\IEEEauthorblockN{1\textsuperscript{st} Francesco Scala}
\IEEEauthorblockA{\textit{Dipartimento di Fisica} \\
\textit{Università di Pavia}\\
Pavia, Italy \\
francesco.scala01@ateneopv.it}
\and
\IEEEauthorblockN{2\textsuperscript{nd} Stefano Mangini}
\IEEEauthorblockA{\textit{Dipartimento di Fisica} \\
\textit{Università di Pavia}\\
Pavia, Italy \\
stefano.mangini01@ateneopv.it}
\and
\IEEEauthorblockN{3\textsuperscript{rd} Chiara Macchiavello}
\IEEEauthorblockA{\textit{Dipartimento di Fisica} \\
\textit{Università di Pavia}\\
Pavia, Italy \\
chiara.macchiavello@unipv.it}
\and
\IEEEauthorblockN{4\textsuperscript{th} Daniele Bajoni}
\IEEEauthorblockA{\textit{Dipartimento di Ingegneria Industriale e dell’Informazione} \\
\textit{Università di Pavia}\\
Pavia, Italy \\
daniele.bajoni@unipv.it}
\and
\IEEEauthorblockN{5\textsuperscript{th} Dario Gerace}
\IEEEauthorblockA{\textit{Dipartimento di Fisica} \\
\textit{Università di Pavia}\\
Pavia, Italy \\
dario.gerace@unipv.it}}

\maketitle

\begin{abstract}
Several proposals have been recently introduced to implement Quantum Machine Learning (QML) algorithms for the analysis of classical data sets employing variational learning means. There has been, however, a limited amount of work on the characterization and analysis of quantum data by means of these techniques, so far. This work focuses on one such ambitious goal, namely the potential implementation of quantum algorithms allowing to properly classify quantum states defined over a single register of $n$ qubits, based on their degree of entanglement. This is a notoriously hard task to be performed on classical hardware, due to the exponential scaling of the corresponding Hilbert space as $2^n$. We exploit the notion of ``entanglement witness'', \textit{i.e.}, an operator whose expectation values allow to identify certain specific states as entangled. More in detail, we made use of Quantum Neural Networks (QNNs) in order to successfully \textit{learn} how to reproduce the action of an entanglement witness. This work may pave the way to an efficient combination of QML algorithms and quantum information protocols, possibly outperforming classical approaches to analyse quantum data. All these topics are discussed and properly demonstrated through a simulation of the related quantum circuit model.
\end{abstract}
\section{Introduction}
\label{sec:intro}
The implementation of machine learning (ML) algorithms on quantum computing hardware, \textit{i.e.} Quantum Machine Learning (QML)~\cite{Dunjko2020}, has established as an independent research field from Quantum Computing (QC) in recent years~\cite{lloyd2013quantum, book_schuld}. In analogy with QC, one of the main goals pursued by QML is to obtain quantum advantage, or \textit{quantum supremacy}~\cite{supremacy2019,PhysRevLett.127.180502,PhysRevLett.127.180501}, with respect to classical machine learning techniques~\cite{Abbas_2021}, exploiting intrinsically quantum resources such as  entanglement. Besides representing a resource for QML, entanglement is also a property of quantum states that is classically difficult to measure, and in this respect QML algorithms could offer an advantage over costly measurement procedures.

Variational Quantum Algorithms (VQAs) are a class of QML models having great potential and currently pursued in several applications~\cite{review_mangini}. They are hybrid protocols whose objective is to find the optimal parameters of a Parametrized Quantum Circuit (PQC)~\cite{Peruzzo_2014,Benedetti2019,Variational,katabarwa2021connecting,PQC_Cerezo_2021,Sim2019, Hubregtsen2021}, \textit{i.e.}, a quantum circuit depending on some classically adjusted parameters, in order to fulfil a given task. VQAs are considered hybrid approaches since, after the three main steps that are performed as quantum tasks (\textit{i.e.}, data encoding, execution of a variational ansatz, and final measurements), the parameters update is performed through classical optimization procedures aimed at minimizing a proper cost function defined for the given problem. This class of algorithms is the closest to classical Artificial Neural Networks (ANNs) trained through backpropagation, and the key elements are usually defined as variational ansatzes containing multiple repetitions of self-similar layers of operations. Due to their similarity to ANNs, PQCs are often referred to as Quantum Neural Networks (QNNs)~\cite{review_mangini}. These models are particularly appealing since they are expected to be practically performing also on currently available quantum hardware.
However, trainability issues, also known as \textit{barren plateaus}, could hinder the efficiency of learning processes~\cite{mcclean_barren, holmes2021connecting, Cerezo_2021, arrasmith2020effect, Patti_2021, arrasmith2021equivalence, marrero2021entanglement}.

QML techniques, and in particular VQAs, have been extensively proposed to analyse large amounts of classical data~\cite{Rebentrost2014} in many different fields such as, for instance, finance~\cite{Egger2020,Martin2021,pistoia2021quantum} and image processing~\cite{neuron,NN,Mengoni2021,rudolph2020generation,Grant2018}. On the other hand, there are very few cases in which QML has been applied to successfully process and classify quantum data~\cite{Grant2018,Cong2019,Yu2019,benedetti2019generative,beer2020, NTangledDataset}, after initial investigations on the auto-analysis of a quantum system own degree of entanglement~\cite{Behrman}. Indeed, the latter promises to be one of the most relevant  applications, owing to the inherent quantum nature of the algorithm itself. As an example, a PQC could be exploited to analyse better and faster an input quantum state, without the need to go through the resource intensive process of quantum state tomography. 

Entanglement is a key feature of multiple quantum information protocols~\cite{Bru__2002, TERHAL2002313, Plenio_2007, Review_ent, appunti_macchia}, such as dense coding, quantum teleportation, and quantum cryptography. Hence, the need for an efficient means of classifying entanglement has also been highlighted in some recent works, where classical ML techniques have been employed~\cite{PhysRevA.98.012315, zhu2021machine, S__2021, Chen_2021, Roik_2021}.
In this work we address the use of QML techniques for entanglement detection. In particular, we analyse a class of quantum states widely employed in quantum computing, the \textit{hypergraph states}~\cite{Rossi2013,Ghio2017}, exploiting the notion of \textit{entanglement witness} \cite{HORODECKI19961,TERHAL2000319,GUHNE20091,Schmidt}. We show how to implement an entanglement witness and how to learn a witness thanks to the use of QNNs. The final outcome of this work is a quantum algorithm able to classify the entanglement of the input state, without any need to perform a measurement on the qubits encoding the state itself. With further improvements this result may pave the way to use of more general entangled states in quantum information protocols.

\section{Theoretical background and Methods}
\label{sec:background_methods}

We hereby address the detection of \textit{full multipartite} entanglement in hypergraph states by using QML techniques. We introduce the basic notions to allow the interpretation and understanding of results shown in the next section.

\subsection{Hypergraph states}
\label{sec:hypergraph_states}

The starting point is the definition of the class of hypergraph quantum states~\cite{Rossi2013}. 
%
One important feature of this class of quantum states is that it coincides with the set of Real Equally Weighted (REW) pure states which can be synthesized with the following state expression, when considering an $n$ qubits state:
\begin{equation*}
    \ket{\psi_f} = \frac{1}{\sqrt{2^{n}}}\sum_{x=0}^{2^n-1}(-1)^{f(x)}\ket{x} 
\end{equation*}
where $\ket{x}$ are the computational basis states, while $f(x)$ is a Boolean function $f :\{0,1\}^n \rightarrow \{0, 1\}$. 
The total number of different possible REW states is $2^{2^n}$, since the action of the function $f$ uniquely defines each state determining which sign goes in front of each component of the computational basis. This tells us that even the hypergraph states set has a dimension of $2^{2^n}$.
The procedure to prepare a hypergraph state makes use of a \textit{hypergraph states generation subroutine} (HSGS), first introduced in~\cite{Rossi2013}, and later used in Python code to run a quantum perceptron on quantum hardware~\cite{neuron}.

The Hilbert space dimension of REW states grows exponentially with the number of qubits, which implies that it is composed of either states that are entangled at different levels or fully separable states. We focus on the case of 3-qubits REW states since in that case the set of states is large enough to have different degrees of entanglement, but its dimension allows us to run the algorithm on quantum computing simulators.

\subsection{Multipartite entanglement}
\label{sec:multip_ent}
The goal of this work is to investigate \textit{multipartite} entanglement, \textit{i.e.}, entanglement shared between more than two qubits, defined as~\cite{appunti_macchia}:

\begin{definition}
An $n$ parties pure state $\ket{\psi}\in\mathcal{H}=\mathcal{H}_A\otimes\mathcal{H}_B\otimes...\otimes \mathcal{H}_N$ is \emph{fully separable} iff it can be written in the following form
$$\ket{\psi} = \ket{a} \otimes \ket{b} \otimes . . . \otimes \ket{n} \quad. $$
It is \emph{k-separable} with respect to a specific partition, iff
$$\ket{\psi} = \ket{\alpha} \otimes \ket{\beta} \otimes . . . \otimes \ket{k} \quad,$$
where we have $k$ subsystems with dimensions $d_\alpha,d_\beta,...,d_k$.
It is \emph{bi-separable} with respect to a specific partition, iff $k=2$, \textit{i.e.}
$$\ket{\psi}= \ket{A} \otimes \ket{B} \quad,$$
where $\ket{A}$ is an $m$ parties state and $\ket{B}$ is an $n-m$ parties state. The state $\ket{\psi}$ is called \emph{fully entangled} (or n-party entangled) if and only if it is not bi-separable with respect to \textit{any} bipartition.
\end{definition}

We now give the definition of the specific measure of multipartite entanglement that we are going to use in the following to classify the hypergraph states~\cite{Ghio2017}:

\begin{definition}
Let $\ket{\psi_n}\in \mathcal{H}_n$ be the pure state of a composite quantum system composed of $n$ subsystems. Let $AB$ be a possible bipartition of the $n$ subsystems with $A = {1, 2, ... , k}$ and $B = {k + 1, ... , n}$ for some $1 \le k < n$. We define the measure of multipartite entanglement of the state as:
\begin{equation}
    \begin{split}
    E(\ket{\psi_n}):&= \min_{AB} E^{AB}(\ket{\psi_n})=\\ 
    &= 1 - \max_{\ket{\phi }_A \ket{\phi}_B, AB} |\bra{\phi}_A\braket{\phi|_B \psi_n}|^2 =\\
    &= 1 - \alpha(\ket{\psi_n})
    \end{split}
    \label{eq:mult_ent}
\end{equation}
where the $\min$ is taken with respect to all possible bipartitions $AB$, the $\max$ is taken over all the possible biseparable states $\ket{\phi_k}_A,\ket{\phi_{n-k}}_B$. 
\end{definition}

The values of the entanglement measure defined in~\eqref{eq:mult_ent} range from 0 (separable state) to 0.5 (maximally entangled state). In our case study with 3 qubits, there are 256 different states with only 64 of them being bi-separable. The remaining 192 are entangled, where 64 have maximal degree of multipartite entanglement and 128 with entanglement equal to 0.25, as can be verified by the direct application of~\eqref{eq:mult_ent} to the whole set of REW states of 3 qubits.

\subsection{Entanglement witness}
\label{sec:ent_wit}

In order to detect entangled states, described in general by a density matrix $\rho$, we exploit the notion of \textit{entanglement witness}, defined as follows~\cite{HORODECKI19961, TERHAL2000319, GUHNE20091}:

\begin{definition}
    An observable, or Hermitian operator, $W$ is called an \emph{entanglement witness} (or witness), if
    \begin{align*}
    &\Tr[W\rho_s] \ge 0 \quad \forall\rho_s\text{ separable,}\\
    &\Tr[W\rho_e] < 0 \quad \text{for at least one entangled $\rho_e$}   
    \end{align*}
holds. Thus, the condition $\Tr[W\rho] < 0$ is sufficient, but not necessary, for the state $\rho$ to be entangled.
\end{definition}

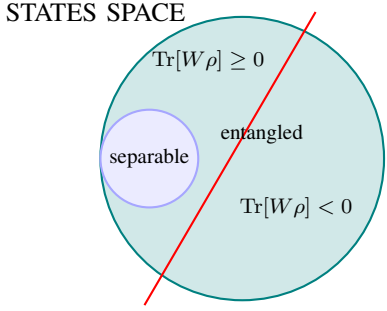
\begin{figure}[]
    \centering
\begin{tikzpicture}[scale=0.65]

\draw [teal, thick, fill=teal!18] (3,3) circle (2.9);
\draw [blue!35, thick, fill=blue!8] (1.1,3) circle (1);

\draw [red, thick](1,0) -- (4.5,6);

\node (tr_neg) at (4.1,2) {\footnotesize $\Tr[W\rho] < 0$};
\node (tr_pos) at (2.3,5) {\footnotesize $\Tr[W\rho] \ge 0$};

\node (sep) at (1.1,3) {\footnotesize separable};
\node (ent) at (3.4,3.5) {\footnotesize entangled};

\node (states) at (0,6) {STATES SPACE};

\end{tikzpicture}
    \caption{Conceptual representation of the convex hypergraph states space, highlighting the distinction between entangled and separable states, and defining the hyperplane (red) determined by the entanglement witness. On the right side of the hyperplane, the witness selects the states that it recognises as entangled (\textit{i.e.}, $\Tr[W\rho] < 0$), on the left side all the others ($\Tr[W\rho] \ge 0$).}
    \label{fig:witness}
\end{figure}

The definition of the entanglement witness has a very specific geometrical meaning that also gives an intuitive notion of what it does. As a matter of fact, the set of states $\rho$ such that $\Tr[W\rho]=0$ defines an hyperplane in the states space. This hyperplane divides the space in two parts: the first, in which $\Tr[W\rho] < 0$, contains only entangled states, while in the other, where $\Tr[W\rho] \ge 0$, lies the whole set of separable states together with some entangled states, see Fig.~\ref{fig:witness}. We say that the entangled states that are in the former side are the ones detected by the witness.

In particular, we refer to a specific kind of entanglement witness that uses a reference entangled state to classify as entangled some other states in the neighborhood of the first one~\cite{GUHNE20091}. The intuition behind this type of witness is that states close to an entangled state must also be entangled. This class is then called \textit{projective}, since it is made of operators in the form:
\begin{equation}
    W = \alpha (\ket {H}) \mathbb {I} - \ket {H} \bra {H}
\end{equation} 
where $\ket {H} \bra {H}$ is the projector over the state of reference, $\ket{H}$, and $\alpha (\ket {H})$ is the same coefficient that we have in~\eqref{eq:mult_ent}. By the linearity of the trace and the fact that $\Tr[\rho]=1$ for every density matrix $\rho$, we can write it as
\begin{equation}
    \begin{split}
    \Tr[\rho W]&=\Tr[\rho(\alpha (\ket {H}) \mathbb {I} - \ket {H} \bra {H})]=\\\
    &=\alpha (\ket {H})- \Tr[\rho \ket {H} \bra {H}] \quad.
    \end{split}
\label{eq:wit_tr}
\end{equation}

Hence, the computation of the entanglement witness is reduced to calculating the overlap between the states $\rho$ and $\ket{H}\bra{H}$, in addition to the calculation of the coefficient $\alpha (\ket {H})$. More specifically, the quantity $\Tr[\rho \ket {H} \bra {H}] = \bra{H}\rho\ket{H}$ defines the fidelity between the states $\ket {H}$ and  $\rho$; if this fidelity exceeds the critical value set by the coefficient $\alpha$, then the expectation value of the witness will be negative and the state $\rho$ must be entangled. 

Our objective in this work is to detect with high fidelity a subset of the entangled hypergraph states defined over 3 qubits. In order to achieve this goal we need to be able to compute $\alpha (\ket {H})$ and the fidelity between an input state and the reference one. For what concerns the calculation of the coefficients, we can avoid to compute the overlap of our states with all the possible biseparable states of a fixed bipartion $AB$ since this is equal to~\cite{Schmidt} 

\begin{equation}
\alpha^{AB} (\ket {\psi})=\max_{k=1,...,R}s_k^{AB}(\ket{\psi})^2
\label{eq:alpha}
\end{equation}

where $s_k^{AB}$ are the Schmidt coefficients, and $R$ is the Schmidt rank. On the other hand, the calculation of the overlap between two REW states can be performed, \textit{e.g.}, by giving an hypergraph state as an input to the Quantum Perceptron presented in~\cite{neuron}, where a fixed hypergraph state of reference is encoded. More in detail, if the input is a state $\ket{\psi_{in}}=U_{i_{in}}\ket{0}^{\otimes n}$, the overlap is calculated by further operating the quantum register with a unitary $U_{w_H}$ corresponding to the reference state $\ket{H}=U_{i_{H}}\ket{0}^{\otimes n}$. Since by definition $U_{w_H} \ket{H}= \ket{1}^{\otimes n}$, $U_{w_H}$ only differs from $U_{i_H}^\dagger$ by an $X^{\otimes n}$ gate and so the second part of \eqref{eq:wit_tr} becomes:
\begin{equation}
\begin{split}
    \Tr[\rho \ket {H} \bra {H}] &=  \Tr[\ket{\psi_{in}}\bra{\psi_{in}} \ket {H} \bra {H}] =|\braket{H|\psi_{in}}|^2 = \\
    & = 
    | \bra{0}^{\otimes n} U_{i_H}^\dagger U_{i_{in}}\ket{0}^{\otimes n}|^2 = \\
    &=| \bra{1}^{\otimes n}U_{w_H} U_{i_{in}}\ket{0}^{\otimes n}|^2
    \quad.
\end{split}
\end{equation}


We measure the overlap by only measuring an ancillary qubit, where the outcome of the computation is transferred thanks to a multi-controlled $X$ gate. In the end, the classification is performed by checking whether the value of the overlap does exceed the threshold set by $\alpha(\ket{H})$ or not. 

\section{Variational learning of an entanglement witness}
\label{sec:var_learn_wit}

Here we briefly outline how to implement a projective entanglement witness using a single artificial neuron simulated through a quantum circuit model. After introducing how to exactly determine an entanglement witness, we now define a procedure on how to approximate its action with the use of a variational quantum perceptron obtained with a PQC~\cite{Variational}. 
The first and most straightforward way is to train the variational model so as to maximize the fidelity with the state used to implement the desired projective witness, just following the steps reported in~\cite{Variational}. On one hand, this PQC could be used as a witness itself since it perfectly mimics the action of a $U_{w_H}$, on the other hand we are not satisfied with the assigned learning task since it does not force the algorithm to try to ``understand'' what entanglement is. More in detail, we would like to give the VQA a supervised learning task and see if it is able to discriminate between entangled and separable states, just like an entanglement witness does. In order to do this, we first implement a \textit{known} witness giving at the input states labeled with 1 only if they are recognised as entangled from the ``real'' witness and 0 otherwise. Then, we want to check if VQAs are able to outperform the witnesses seen so far  by detecting more (or at least more maximally entangled) states. To fulfil this goal we label the states with 1 if they are entangled and 0 if they are separable, meaning that we want to learn an \textit{unknown} witness.

\begin{figure*}[]
    \centering
    \begin{tikzpicture}
    \node[scale=0.9]{
    \begin{quantikz}
        \lstick{$\ket{0}$} & \gate[1]{\text{$H$}} &\gate[1]{\text{$Z$}} & \gate[1]{\text{$R_y(\theta_0)$}}\gategroup[3,steps=11,style={dashed,
rounded corners,fill=blue!20, inner xsep=2pt}, background]{{\sc PQC}} & \ctrl{1}&\ctrl{2}&\qw &\qw & \gate[1]{\text{$R_y(\theta_3)$}} & \ctrl{1}&\ctrl{2}&\qw &\qw & \gate[1]{\text{$R_y(\theta_6)$}}&\qw & \ctrl{1}\\
        \lstick{$\ket{0}$} &\gate[1]{\text{$H$}} &\qw & \gate[1]{\text{$R_y(\theta_1)$}}&  \targ{}   &\qw           & \ctrl{1} &\qw & \gate[1]{\text{$R_y(\theta_4)$}}&  \targ{}   &\qw           & \ctrl{1} &\qw & \gate[1]{\text{$R_y(\theta_7)$}}&\qw & \ctrl{1}\\
        \lstick{$\ket{0}$} &\gate[1]{\text{$H$}} &\qw & \gate[1]{\text{$R_y(\theta_2)$}}&  \qw   &\targ{}         & \targ{} &\qw & \gate[1]{\text{$R_y(\theta_5)$}}&  \qw   &\targ{}         & \targ{} &\qw & \gate[1]{\text{$R_y(\theta_8)$}}&\qw & \ctrl{1}\\
        \lstick{$\ket{0}_a$} &\qw &\qw &\qw &\qw& \qw &\qw &\qw &\qw &\qw &\qw& \qw &\qw &\qw &\qw & \targ{} & \meter{}
    \end{quantikz}};
    \end{tikzpicture}
        \caption{Quantum circuit representation of the encoding of the string [0, 0, 0, 0, 1, 1, 1, 1] followed by our variational ansatz made of $CNOT$ gates and $R_y$ rotations initialized with random angles $\theta _i$. The result of the computation is then transferred to the ancilla qubit thanks to a multi-controlled $NOT$ gate.}
    \label{fig:kn_wit_circ}
\end{figure*}
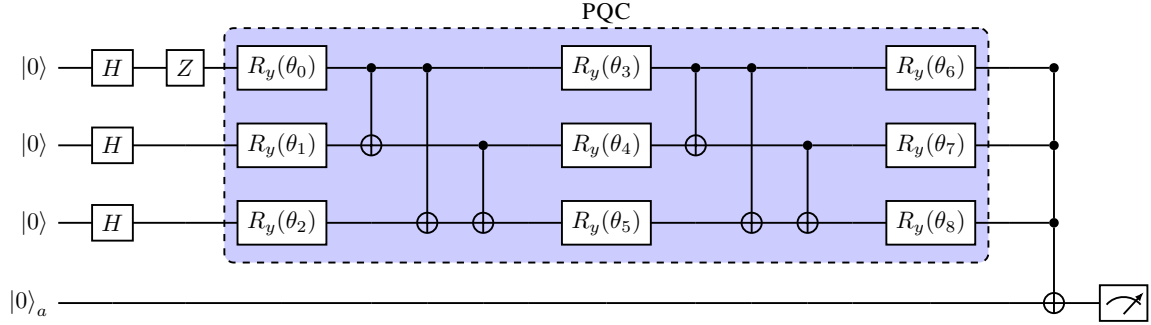

An example of quantum circuit implementing the state encoding and our variational model is shown in Fig.~\ref{fig:kn_wit_circ}.
We can retrieve if a state is on one side of the hyperplane or the other by measuring the ancillary qubit. We check whenever the ancilla is in state $\ket{1}$, meaning that the state of the quantum register of interest is $\ket{111}$. The ratio between the number of times that this happens and the total number of times we execute the circuit gives an estimate of $\Tr[\rho \ket{V(\bm{\theta})}\bra{V(\bm{\theta})}]$, where $V(\bm{\theta})$ is the unitary operation performed by the variational part of the circuit.

In order to reach the optimal values of the parameters defining the PQC, VQAs need a cost function to be minimized through the use of an optimizer. We choose COBYLA~\cite{cobyla} as our classical optimizer. This is a gradient-free optimization algorithm that creates a region with $N+1$ vertices (polytope) in the parameters space on which the cost function is evaluated, $N$ being the number of variational parameters. The optimal point is found by progressively reducing the size of the region. 

\begin{figure}[t]
    \centering
\begin{tikzpicture}[scale=0.65]

\draw [teal, thick, fill=teal!18] (3,3) circle (2.9);
\draw [blue!35, thick, fill=blue!8] (1.1,3) circle (1);

\draw [orange, fill=orange!25] (0.6,2.7) circle (0.1);
\draw [orange, fill=orange!25] (1.8,3) circle (0.1);
\draw [orange, fill=orange!25] (1.3,2.2) circle (0.1);

\draw [violet, fill=violet!25](2.7,5) circle (0.1);
\draw [violet, fill=violet!25] (1,1.8) circle (0.1);
\draw [violet, fill=violet!25] (3.7,4.9) circle (0.1);
\draw [violet, fill=violet!25] (1.5,4.5) circle (0.1);
\draw [violet, fill=violet!25] (3.7,4.1) circle (0.1);
\draw [violet, fill=violet!25] (4.7,3.1) circle (0.1);
\draw [violet, fill=violet!25] (3.8, 1.7) circle (0.1);
\draw [violet, fill=violet!25] (1.9,1.1) circle (0.1);

\draw [red, thick](1,0) -- (4.5,6);
\draw [red,thick](3.5,-0.6) -- (7,5.4);

\node (sep) at (1.1,3) {\footnotesize separable};
\node (ent) at (3.4,3.5) {\footnotesize entangled};

\node (states) at (0,6) {STATES SPACE};
\draw[<-,thick] (3,3) -- (4.9,2.1);
\draw[ thick] (5,5.5) rectangle (7.5,6.5);
\draw [violet, fill=violet!25] (5.2,6.2) circle (0.1);
\node (legend_1) at (6.28,6.2) {\footnotesize entangled};
\draw [orange, fill=orange!25] (5.2,5.775) circle (0.1);
\node (legend_2) at (6.25,5.75) {\footnotesize separable};

\end{tikzpicture}
    \caption{Schematic representation of the convex states space, highlighting the distinction between entangled (violet) and separable (orange) states, with the hyperplane (red) defined by the entanglement witness. On the right side of the hyperplane, the witness leaves the states that it recognises as entangled, on the left side all the others. The learning goal is to move our hyperplane as close as possible to the convex set of separable states without crossing it, thus leaving on the other side as many entangled states as possible.}
    \label{fig:moving_hyperplane}
\end{figure}
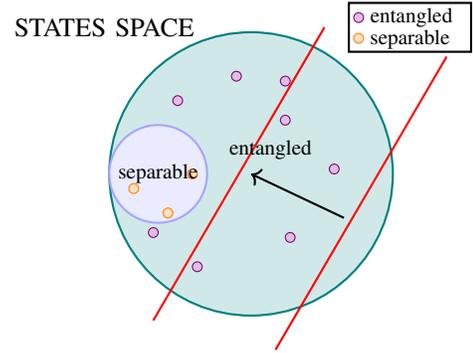

The choice for a good cost function is crucial to achieve optimal performances. The classification performed by an entanglement witness is non trivial because it defines a hyperplane that leaves on one side some entangled states and on the other side \textit{all} the separable states together with the remaining entangled states. In this light, our goal is to classify states with respect to their degree of entanglement, without making \textit{any} mistake on separable states, while trying to maximize the number of recognised entangled states. {Here we stress that the requirement to perfectly classify the separable states is stronger than the need of maximizing the detected entangled one.} In other words, we can say that the learning goal is to move our hyperplane as close as possible to the convex set of separable states without crossing it and leaving on the other side as many entangled states as we can. This is schematically summarized in Fig.~\ref{fig:moving_hyperplane}. 

In order to fully understand what we just introduced, we need to present two different quantities that will be fundamental in the following: \textit{Precision} and \textit{Recall}. Precision indicates the portion of true positive with respect to all the data classified as positive, for this reason it is also called Positive Predictive Value (PPV), while the Recall is the ratio between true positives and the number of data belonging to the positive class, and it is also called True Positive Rate (TPR) or sensitivity~\cite{powers2020evaluation}:
\begin{equation}
  \text{Precision} = \frac{tp}{tp + fp} \quad \quad, \quad \quad
     \text{Recall} = \frac{tp}{tp + fn} \quad.
\end{equation}

Applying these concepts to our problem, a Precision value of 1 means that we are correctly classifying as entangled at least one state, and still not crossing the set of separable states; while maximizing the Recall allows closing the gap between the hyperplane and that set. Since we cannot allow \textit{any} cut of the separable states set by the hyperplane, while we can accept to have fewer detected entangled states, we would like to assign different weights to Precision and Recall so as to give some priority to the first one of the two. This requirement is perfectly fulfilled by the \textit{$F_\beta$-score}~\cite{powers2020evaluation}:
\begin{equation}
    F_\beta = (1 + \beta^2) \cdot \frac{\mathrm{Precision} \cdot \mathrm{Recall} }{ \beta^2 \cdot \mathrm{Precision} + \mathrm{Recall}}
\end{equation}
where $\beta$ is chosen such that Recall is considered $\beta$-times as important as Precision. Since we want to put emphasis on Precision we will have $0<\beta<1$.

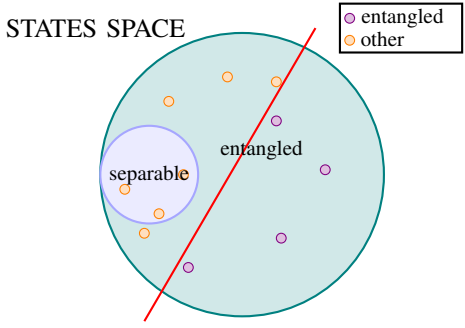
\begin{figure}[t]
    \centering
    \begin{tikzpicture}[scale=0.65]
    
    \draw [teal, thick, fill=teal!18] (3,3) circle (2.9);
    \draw [blue!35, thick, fill=blue!8] (1.1,3) circle (1);

    \draw [orange, fill=orange!25] (1.8,3) circle (0.1);
    \draw [orange, fill=orange!25] (1.3,2.2) circle (0.1);
    \draw [orange, fill=orange!25] (0.6,2.7) circle (0.1);
    \draw [orange, fill=orange!25] (2.7,5) circle (0.1);
    \draw [orange, fill=orange!25] (1,1.8) circle (0.1);
    \draw [orange, fill=orange!25] (3.7,4.9) circle (0.1);
    \draw [orange, fill=orange!25] (1.5,4.5) circle (0.1);
    
    \draw [violet, fill=violet!25] (3.7,4.1) circle (0.1);
    \draw [violet, fill=violet!25] (4.7,3.1) circle (0.1);
    \draw [violet, fill=violet!25] (3.8, 1.7) circle (0.1);
    \draw [violet, fill=violet!25] (1.9,1.1) circle (0.1);
    
    \draw [red, thick](1,0) -- (4.5,6);    
    \draw[ thick] (5,5.5) rectangle (7.5,6.5);
    \draw [violet, fill=violet!25] (5.2,6.2) circle (0.1);
    \node (states) at (0,6) {STATES SPACE};
    
    \node (sep) at (1.1,3) {\footnotesize separable};
    \node (ent) at (3.4,3.5) {\footnotesize entangled};
    \node (legend_1) at (6.28,6.2) {\footnotesize entangled};
    \draw [orange, fill=orange!25] (5.2,5.75) circle (0.1);
    \node (legend_2) at (5.87,5.775) {\footnotesize other};
    \end{tikzpicture}
    \caption{The picture shows a simplified chart in which the states had been colored with violet if they must be recognised as entangled by the witness and colored with orange if they must be classified as ``other''. The second class contains both entangled and separable states.}
    \label{fig:known_wit_repr}
\end{figure}

The task of reproducing a \textit{known} witness should be achievable in a relatively simple way, since our problem is to find a hyperplane that divides the points labeled as entangled from the ``other'', as represented in Fig.~\ref{fig:known_wit_repr}. In other words, the classification problem is a linear one in the states space and we already know that a solution does exist.

This said, in this particular case our goal can be reduced to maximize the number of correctly classified states, as commonly done in ML problems. So, instead of the $F_\beta$-score we choose a widespread cost function, the \textit{cross entropy}. The latter is a measure of the divergence between the labels ($y_i$) and the predicted probabilities ($p_i$),
\begin{equation}
   cross \ entropy = - \sum _{i=0}^{m-1}y_i\log p_i + (1-y_i)\log (1-p_i)
\end{equation}
where $m$ is the number of data samples.
We leave the use of the $F_\beta$-score as a cost function for the learning of an \textit{unknown} witness, given that it encodes the problem of optimizing the position of the hyperplane. 

The values of the parameters for the learning process are randomly initialized. Being aware of how the optimizer operates and of the fact that the optimization problem we are facing is full of local minima, we decided to iterate the algorithm with random starting points or with a local minimum as starting point, whenever an interesting one is found. {We define as ``interesting'' points those that return values of the cost below a certain threshold.} This intuition is dictated by the fact that, most often, local minima could be close to each other in the parameters space, and one expects that the global ones would be as well.

\section{Results and discussion}
\label{sec:results}

Here we provide the results obtained by numerical simulations of the PQCs introduced in the previous section. In particular, we compare the learning of entanglement witness with exact results, showing very good performances of our hybrid quantum/classical algorithm. All these results were obtained by using the \textit{qasm\_simulator} of Qiskit~\cite{Qiskit}.

\subsection{Exact witness computation}
\label{sec:exact_wit}

Among all the 3-qubits hypergraph entangled states we choose a state of reference that has the maximum degree of entanglement. This particular choice is aimed at pursuing the detection of entangled states with a high degree of entanglement. Once the state is selected, as a first step we can compute $\alpha(\ket{H})$, and then perform the computation of $\Tr[\rho \ket{H}\bra{H}]$ with the quantum perceptron for every possible REW state, $\rho$.

\begin{figure}[t]
    \centering
    \includegraphics[width=\columnwidth]{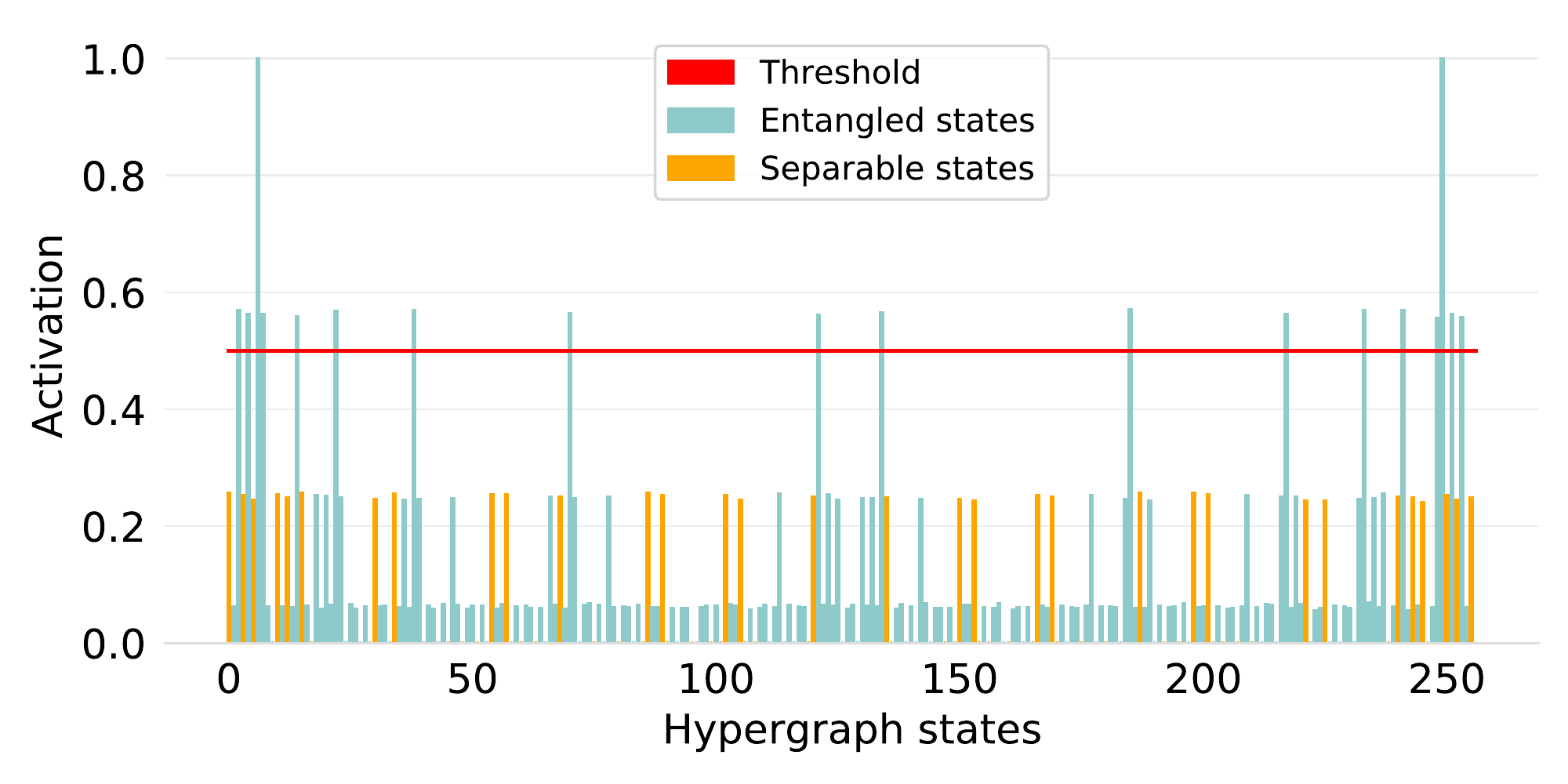}
    \caption{
    Bar chart showing the quantum perceptron activation, with respect to the threshold equal to $\alpha(\ket{H})$, for both the entangled (light blue) and the separable (orange) states. These results are obtained when taking the maximally entangled reference state corresponding to the string [0, 0, 0, 0, 0, 1, 1, 0].  
    }
    \label{fig:wit_detect}
\end{figure}

The bar chart in Fig.~\ref{fig:wit_detect} shows the activations of the quantum neuron implementing the witness with threshold set by $\alpha(\ket{H})$ for a fixed maximally entangled state of reference $\ket{H}$ of 3 qubits. The witness efficiently recognises 18 states as entangled out of the 192 3-qubits hypergraph entangled states. The two highest spikes correspond to the states $\ket{H}$ itself and the state that differs from $\ket{H}$ by an overall phase of $e^{i\pi}$. {This is due to the fact that the perceptron activation is invariant to an overall sign change~\cite{neuron}. As a consequence, the activation scheme is symmetric. We will call states differing  from a global minus sign ``complementary''.} 

Apart from the state $\ket{H}$ itself and its complementary counterpart, all the other states found have a degree of entanglement equal to 0.25. This happens independently of the particular maximally entangled state of reference. Hence, in our case the exact witness does not to identify maximally entangled states, except for the reference one.

\subsection{Learning of a known witness}
After establishing that an exact computation of the projective entanglement witness is possible, we would like to approximate its action by using a VQA with the ansatz presented in Sec.~\ref{sec:var_learn_wit}. In essence, we aim at performing a supervised QML task allowing to learn a known witness, meaning that the VQA is able to classify certain selected hypergraph states as entangled.


The model is trained using a set of reduced dimension in order to have manageable learning times, since we are working with classical computers simulating the behaviour of truly quantum systems. In particular, we include all the states that must be certainly classified as entangled by the witness (labeled with 1), then we take half of the remaining entangled states (which must lie on the other side of the hyperplane defined by the witness),\footnote{Here we use only half of the states since we know that the perceptron activates in the same way for states differing  from a global minus sign.} and we label with 0 a subset of 60\% of these and a  further half of the separable states. It is important to point out that we shuffle the data so as to avoid any possible learning pattern in the general structure of the data set.

Once the states set is ready, we need to fix the hyperparameters of the variational model. The first one is the number of layers of the circuit, which we have fixed to 2, in order to reduce the complexity and try to avoid the occurrence of barren plateaus~\cite{mcclean_barren}. We set the threshold value to $0.5$, since we want to emulate the classification performed by a projective entanglement witness with a maximally entangled state as a reference. If the threshold is exceeded by the activation value of our PQC, then the state is certainly entangled, otherwise we do not know.


As expected, for all the 64 possible maximally entangled states of reference the VQA learns optimal values for the parameters, enabling the PQC to perform an entanglement detection of all the 18 entangled states with activation values of the quantum perceptron close to the ones provided by the true witness (see Fig.~\ref{fig:wit_detect}). More in detail, one finds that the average cross entropy, between the exact activations and the ones obtained with the learnt witness, is equal to $0.26926 \pm 5.8\cdot10^{-4}$.

\subsection{Learning of an unknown witness}

Our ultimate goal is now to explore the full learning capabilities of the VQA. More in detail, we would like to see whether or not the PQC can implement an unknown entanglement witness that can (in some way) perform a better classification than the previous one. For this reason, here we build a witness without giving any extra information about which entangled states have to be recognised. We assign label 1 to all the entangled states and label 0 to the separable ones. Then, we take the $60\%$ of the entangled states and $90\%$ of the separable ones as a training set\footnote{Even in this case, to reduce the computational cost of the learning, we use only half of the states exploiting the symmetry invariance. Then the test set contains also all the complementary states of the training set.}. 

As anticipated in the previous section, we will make use of the $F_\beta$-score in order to construct a cost function that is ``problem-inspired''. As a matter of fact, given that we want to learn an unknown witness, using as a cost function the cross entropy could be dangerous since it does not favor Precision over Recall and this in principle could lead our witness to classify a separable state as entangled, which is exactly what we do not want. 
Since we would rather have a minimization problem, we take as a cost function $C=1-F_\beta$. The last hyperparameter to be fixed is $\beta$, and we choose it to be $1/30$. This allows to have very low values of the cost function whenever we have Precision equal to 1 whatever the value of Recall is and far higher values of the cost function if Precision is less than 1. Once we have Precision equal to one the small changes of value of the cost function due to the Recall still allows the minimization.
%

\begin{table}[]
\centering
\caption{Unknown witness metrics for train and test in our case study.}
\begin{tabular}{c|c|c|c|}
\cline{2-4}
\textbf{}                            &  \bm{$F_{\beta}$} & \textbf{Precision} & \textbf{Recall} \\ \hline

\multicolumn{1}{|c|}{\textbf{Train}} & 0.9907    & 1.0       &     0.1053     \\ \hline

\multicolumn{1}{|c|}{\textbf{Test}}  &              0.9595     &      1.0             &       0.0256          \\ \hline
\end{tabular}
\label{tab:metrics}
\end{table}

\begin{figure}[]
\centering
        \includegraphics[width=\columnwidth]{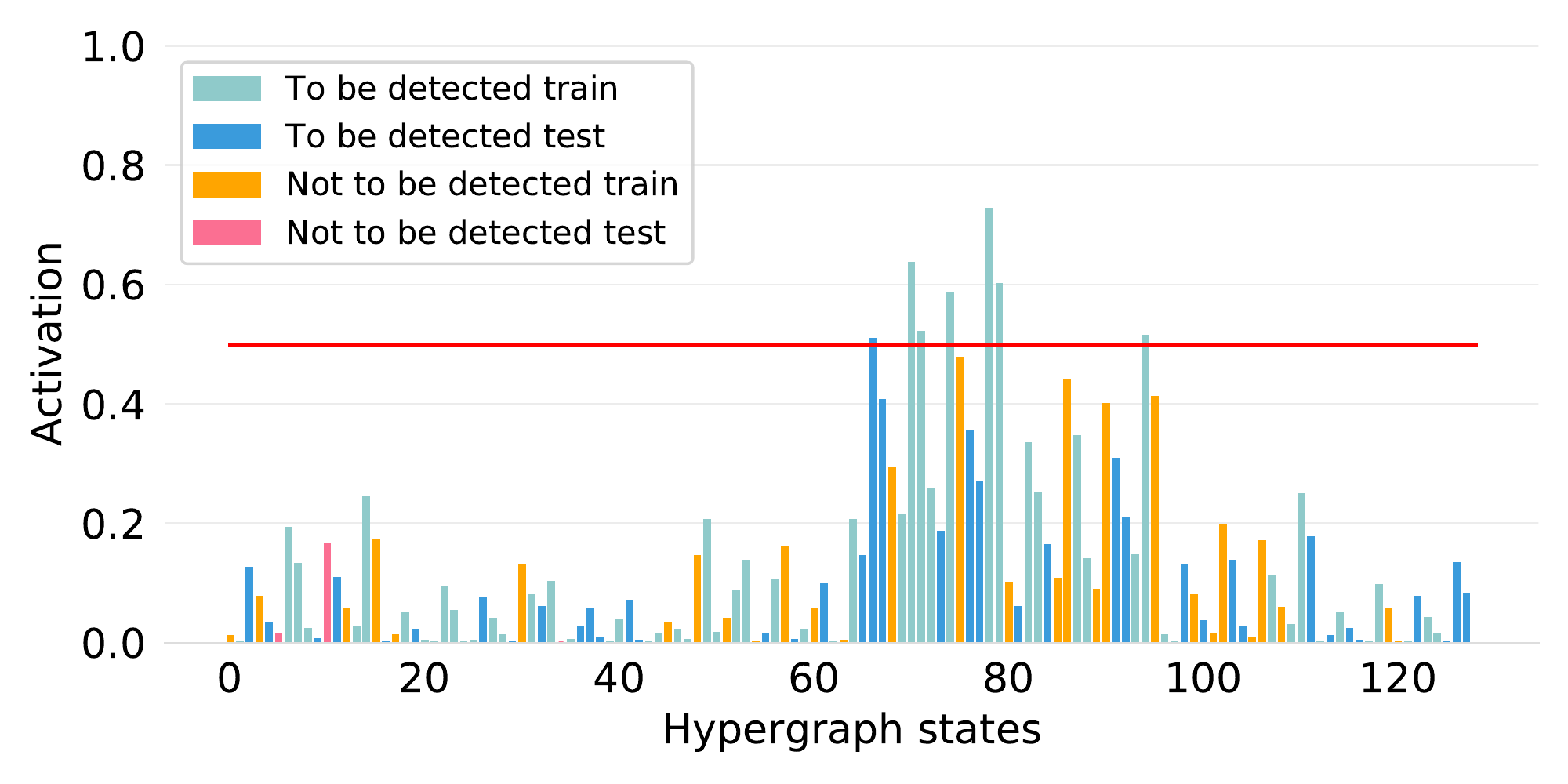}
        \caption{{Bar chart showing the activations of the variational perceptron, with the found optimal parameters allowing to implement an unknown witness, with respect to the threshold equal to $0.5$. Only half of the states is shown, since the activations are symmetric.}}
        \label{fig:var_un_witness_detect}
\end{figure}

After the learning procedure, the PQC is able to efficiently identify entangled hypergraph states of 3 qubits. This is demonstrated by our case study with the fixed random seed, in which 6 entangled states out of the 57 given in the train set are detected and the variational witness does not make any mistake on the separable states, confirming that our choice of the cost function was appropriate. Moreover, our model seems to have generalization capabilities,
since when applied to the test set it is able to detect one more entangled state (and its complementary state), and it is still able to correctly classify separable states (see Tab.~\ref{tab:metrics} and Fig.~\ref{fig:var_un_witness_detect})\footnote{Since the initialization of the parameters is random and the number of iterations can be varied, one may in general obtain slightly different metrics values with respect to the particular case study reported.}.

The most relevant result obtained with this learning procedure is that our variational witness does recognise \textit{maximally entangled} states, which the projective witness built using a reference entangled hypergraph state was not able to do. This is a remarkable achievement since most quantum information protocols need maximally entangled states. More in detail, the PQC identifies a total of six maximally entangled states (two in the training set and one in the test set, plus their complementary counterpart). We would also like to emphasize that there is no activation equal to 1, as one can see from Fig.~\ref{fig:var_un_witness_detect}, meaning that our PQC is implementing a hyperplane corresponding to an unknown reference state that is not part of the REW states set.

Even if the entangled states recognised by the learnt unknown witness are less than the ones detected by the known one, we highlight that our algorithm allows to detect also maximally entangled states whereas the exact witness implementation fails to do that with all 64 possible maximally entangled states of reference.

\section{Conclusions}

There have been few studies in which QML has been employed to analyse quantum data, so far. In most of the previous works addressing entanglement detection with QML, the accuracy levels were limited due to the complexity of the problem~\cite{Grant2018}, highlighting that the application of QML techniques to analyse quantum states is still at an initial stage. 

Here we implement a hybrid quantum-classical algorithm allowing to learn an approximate (known and unknown) projective entanglement witness~\cite{Ghio2017}, through the use of a recently introduced quantum perceptron algorithm~\cite{neuron, Variational}. These results might be relevant for applications of QML algorithms to perform analysis of quantum data in near-term quantum computers and allowing to outperform conventional approaches to solve the same task. 

While classical ML algorithms have recently been employed to learn entanglement witnesses~\cite{PhysRevA.98.012315, zhu2021machine, S__2021, Chen_2021, Roik_2021}, our work provides the first example, to the best of our knowledge, in which an entanglement witness is learnt by purely QML techniques avoiding the requirement of tomographically complete data. As a matter of fact, a tomography (or at least some projections~\cite{Roik_2021}) of the quantum state is required in order to give this one as input to a classical ML algorithm.
In this view, the use of a quantum computer allows to directly work on the input state without any previous tomography or projection.

There are many interesting directions worth exploring to extend the reach of the proposed technique. Future works may analyse more general classes of entangled states~\cite{NTangledDataset}, and comparing the capabilities of different ans\"{a}tzes and optimizers, as well as changing the entanglement measure or modify the cost function with weights to take into account also different entanglement levels. These works will also benefit from the simulation of quantum noise, and the running on real quantum hardware in order to test the robustness of the approach.
\printbibliography

\end{document}